\documentclass[11pt]{article}
\usepackage{graphicx}
\usepackage{amsmath}
\usepackage{amsbsy}
\usepackage{amssymb}

\title{Dynamics of poroelastic filaments}
\author{J. M. Skotheim  \& L. Mahadevan \footnote{current address: Harvard University, Cambridge, MA 02138, USA; {\em Email: lm@deas.harvard.edu}} ~~ \\
Department of Applied Mathematics and Theoretical Physics \\
University of Cambridge \\
 Wilberforce Road, Cambridge CB3 0WA, UK \\  
 \\ 
 {\it to appear in the Proceedings of the Royal Society} \\ \\}
\begin{document}

\maketitle

We investigate the stability and geometrically non-linear dynamics of slender rods made of a linear isotropic poroelastic  material. Dimensional reduction leads to the evolution equation for the shape of the {\it poroelastica} where, in addition 
to the usual terms for the bending of an elastic rod, we find a term that arises from fluid-solid interaction.
Using the {\it poroelastica} equation as a starting point, we consider the load controlled and displacement controlled planar buckling of a slender rod,  as well as the closely related instabilities of a rod subject to twisting moments and compression when embedded in an elastic medium. 
This work has applications to the active and passive mechanics of thin filaments and sheets made from 
gels, plant organs such as stems, roots and leaves, sponges, cartilage layers and bones.



\section{Introduction}

Poroelasticity is the continuum theory used to describe the behaviour of a biphasic material in which
fluid flow is coupled to the elastic deformation of a solid skeleton (see Selvadurai (1996), Wang (2000) and references therein).
The first applications of this theory were to geological problems such as consolidation of saturated soil under a uniform load (Biot 1941).  Since then the theory has grown to cover  many and varied applications, some of which are displayed in Table 1.  

If a medium having interstitial fluid of viscosity $\nu$ is forced to oscillate with a characteristic time $\tau$, the Stokes' length of the motion, $\hbox{L}_{\hbox{s}} = \sqrt{\nu\tau}$, will characterise the range of influence of the solid into the fluid. If $\hbox{L}_{\hbox{s}} \ll l_p$ (the pore length scale) the fluid within the pores moves out of phase relative to the solid. On the other hand, if  $\hbox{L}_{\hbox{s}} \gg l_p$, the fluid will only move relative to the solid when the volume fraction of solid matrix changes locally. This limit ($\hbox{L}_{\hbox{s}}\gg l_p$) was first considered by Biot (1941) for an isotropic poroelastic material. Later work using averaging techniques led to equations of the same form as well an understanding of how the microstructure of the material influences the constitutive equations of the material (Auriault \& Sanchez-Palencia 1977, Burridge \& Keller 1981, Mei \& Auriault 1989, Lydzba \& Shao 2000).

Whereas geological applications are concerned primarily with bulk behaviour, many engineering, physical and biological applications have extreme geometries which allow for the application of asymptotic methods to reduce the dimension of the problem. Some examples  include the active and passive mechanics of thin filaments and sheets made of gels,  plant organs such as stems, roots and leaves,  sponges, cartilage layers, bones etc.  

In this paper we use the constitutive behaviour of a linear isotropic poroelastic solid to investigate the stability and dynamics of slender rods made of this material. In \S \ref{goveq} we give a physically motivated derivation of the constitutive equations for a poroelastic material. In \S \ref{buckling} we use the bulk poroelastic  constitutive equations to determine the equation for the time dependent bending of a slender poroelastic rod subject to an externally applied compressive force $P$.  Dimensional reduction leads to the equation for the {\it poroelastica}, where in addition to the usual terms due to the bending of an elastic rod we find a term due to the fluid resistance. It arises from the fluid-solid interaction and has a form similar to that in a Maxwell fluid (Bird, Armstrong \& Hassager 1987). In \S \ref{Pcontrolled} we solve the problem of load controlled buckling. Although the poroelastic nature of the material does not change the buckling threshold or the final stable shape, it governs the dynamics of the system as it evolves from the unstable to the stable state.  Both the short and the long time limit are investigated investigated using asymptotic methods.  We then use numerical methods to corroborate our asymptotic approaches and follow the non-linear evolution of the poroelastica. In \S \ref{dcb} we treat the problem of displacement controlled buckling and compare the results of the {\it poroelastica} with those of the classical {\it elastica} under similar loading conditions. In \S \ref{twist} we consider the linear instability of a slender poroelastic filament embedded in an infinite elastic medium, when it is subjected to an axial twisting moment  and an axial thrust. Finally, in \S 7, we summarise our results and discuss possible applications to such problems as the mechanics of cartilaginous joints and rapid movements in plants.
\begin{table}
\label{chart}
\begin{center}
\includegraphics[width=10cm]{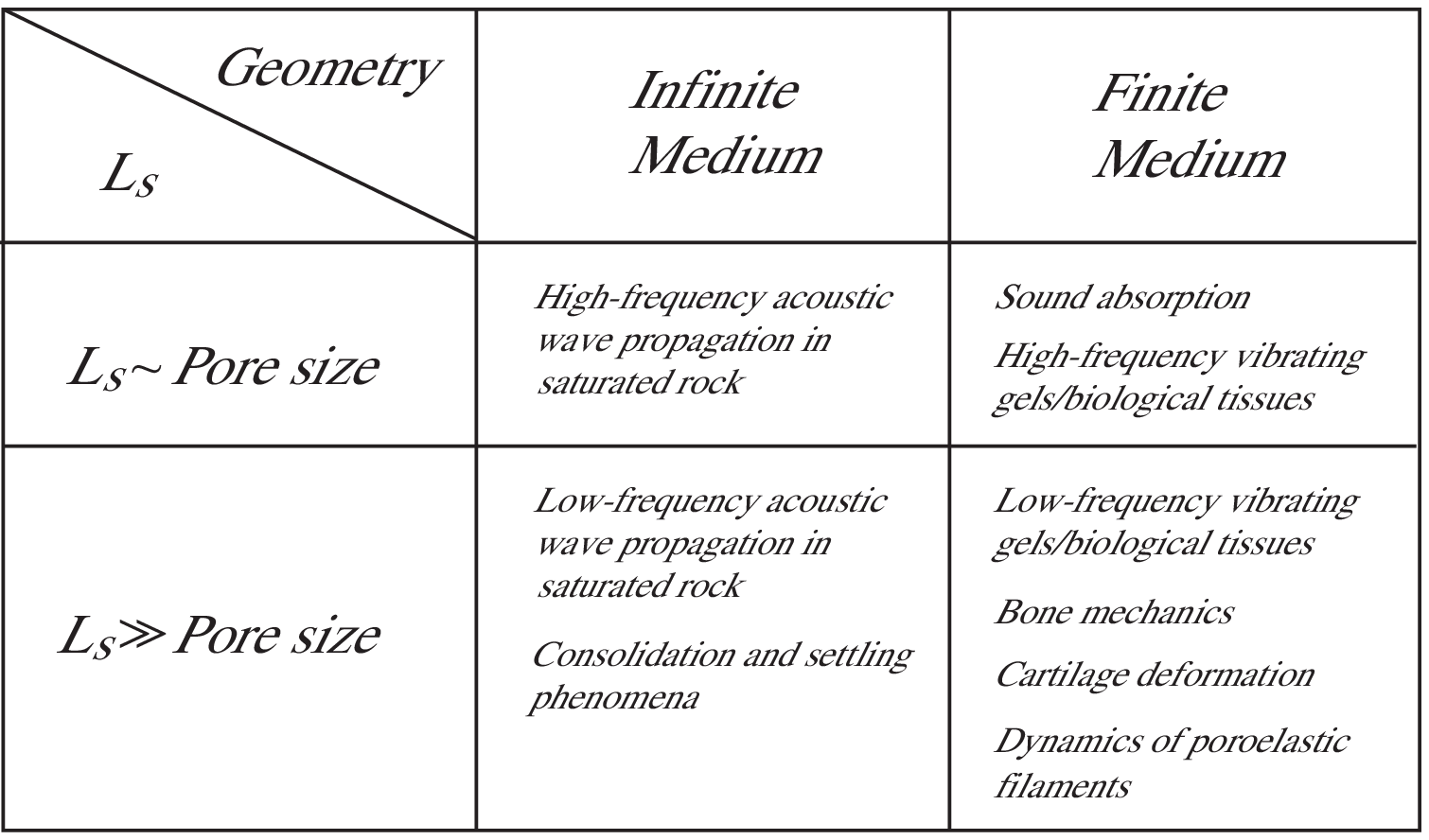}
\par
\caption{
Chart illustrating various applications of poroelasticity.  $\hbox{L}_{\hbox{s}} =  \sqrt{\nu\tau}$ 
is the Stokes' length, where $\nu$ is the kinematic viscosity of 
the interstitial liquid,  and $\tau$ is the time scale of the motion. }
\end{center}
\end{table}

\section{Governing equations for poroelastic media}
\label{goveq}

We begin with the equations for a homogeneous, elastic isotropic poroelastic material in the limit where the
Stokes' length, $\hbox{L}_{\hbox{s}}=\sqrt{\nu\tau} $ is much larger than the system size $l_m$ and further that $l_m  $ is much larger than the pore size
$l_p$. We will also neglect inertial effects in the solid and liquid phases.  In this limit 
the viscous resistance to fluid flow in the pores is balanced by the pressure gradient so that the momentum balance in the fluid yields
\begin{equation}
\label{bala}
\rho\nu\nabla_{l_p}^2{\bf v} - \nabla p -\nabla_{l_p}p_p = 0,
\end{equation}
where ${\bf v}$ is the fluid velocity with characteristic scale $V$, $\rho$ is the fluid density,
$\nabla$ and $\nabla_{l_p}$ denote gradients on the system scale and the pore scale respectively, $p$ is the macroscopic pressure driving the flow, and $p_p$ is the microscopic pressure in the pore. 
When the pore scale and system size are well separated ($l_p/l_m \ll 1$)   (2.1) gives the following scaling relations
\begin{equation}
p \sim \frac{l_m\rho\nu V}{l_p^2} \gg \frac{\rho\nu V}{l_p} \sim p_p.
\end{equation}
Thus the dominant contribution to the fluid stress in the medium arises from the pressure. The simplest stress-strain law for the composite medium then arises  by considering the linear superposition of the dominant components of the fluid and solid stress tensor. Assuming that the elastic behaviour of the solid skeleton is well characterised by Hookean elasticity (i.e. the strains  are small),  we can write the following constitutive equation for the poroelastic medium (see Appendix A for a derivation using the method of multiple scales):
\begin{equation}
\label{CE}
 \boldsymbol{\sigma} = 2\mu{\bf e} + \lambda \nabla \cdot {\bf u}\,{\bf \hbox{I}} 
- \alpha p{\bf \hbox{I}}.
\end{equation}
 Here $\boldsymbol{\sigma}$ is the stress tensor, ${\bf u}$ is the displacement field, ${\bf e} = (\nabla{\bf u} + \nabla{\bf u}^{T})/2$ is the linearised strain, $\mu$ and
$\lambda$ are the effective Lam\'e coefficients of the material (dependent on the material
properties {\it and} the microstructure), $\alpha$ is related to the fluid volume fraction, but includes a contribution from the pressure in the surrounding fluid (see Appendix A), and $\bf I$ is the identity tensor in 3 dimensions.   These material parameters can be derived using microstructural information (see Appendix A).   In the limit when inertia can be neglected, the equations of equilibrium are
\begin{equation}
 \nabla \cdot \boldsymbol{\sigma} =0. 
 \label{F}
\end{equation}
Mass conservation and continuity requires that the rate of dilatation of the solid is balanced by the differential motion between the solid and fluid in a poroelastic solid. This yields (see Appendix A for a derivation using the method of multiple scales)
\begin{equation}
\label{Cont}
\nabla \cdot {\bf k}\cdot\nabla p  = \beta \partial_t p + \alpha\partial_t\nabla\cdot{\bf u} , 
\end{equation}
where the solid skeleton is composed of a material with bulk modulus $\beta^{-1}$
($\ne \lambda +2\mu/3$ since the lam\'e coefficients $\lambda$ and $\mu$ are for the composite
material and take into account the microstructure, while $\beta^{-1}$ is independent of the 
microstructure) 
 and  ${\bf k}$ is the fluid permeability tensor of the solid matrix. In words, (\ref{Cont}) states that
the flux of fluid into a material element is balanced by the change in solid volume due to the bulk compressibility of the matrix.
For a rigid incompressible skeleton, $\beta = 0$. 
We will assume that the solid skeleton is isotropic so that ${\bf k} =k {\bf \hbox{I}} $; however, many structured and biological materials are anisotropic and 
one may need to revisit this assumption.  Equations (\ref{CE}), (\ref{F}) and (\ref{Cont}) when subject to appropriate boundary conditions describe the evolution of displacements $\bf u$ and fluid pressure $p$ in a poroelastic medium. The typical values of the parameters for a soft gel are $\alpha\sim1$, $\mu\sim\lambda\sim10^6Pa$, and $k\sim10^{-12}m^2/sec\,Pa$.

\begin{figure}
\begin{center}
\includegraphics[width=12cm]{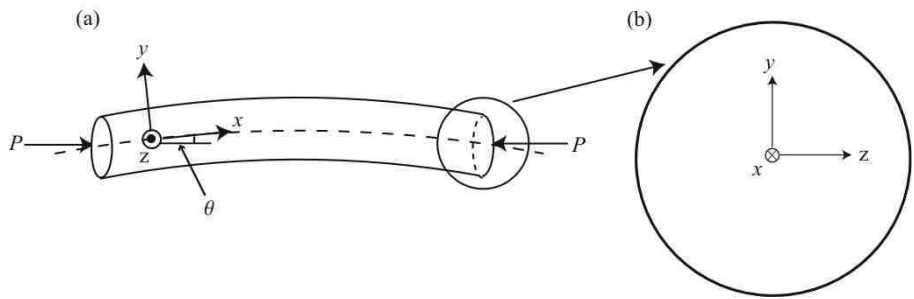}
\par
\caption{Schematic diagram of (a) a bent rod, where $\theta(x)$ is the angle between the deformed and undeformed tangent vector,
$x,y$ and $z$ are body-fixed coordinates in the reference frame of the rod; (b) the circular cross section.}
\label{s}
\end{center}
\end{figure}

\section{Equations of motion for a slender filament}
\label{buckling}

We consider a naturally straight slender circular rod of length $L$ and radius $R \ll L$ with a tangent to the centre line that makes an angle $\theta$  with the horizontal (see Figure \ref{s}).   At its ends an axial  force $P$ is applied suddenly at time $t=0$.  We will further assume that the lateral surfaces of the filament are free of tractions. This assumption could break down when the solid matrix is very dilute, so that interfacial forces become comparable to the internal forces in the filament, but we will not
consider this case here.  The slenderness of the filament implies that the axial stresses vary rapidly across the cross-section and much more slowly along it, so that we can use an averaging procedure to deduce low-dimensional equations that describe the motion of the filament. This long-wavelength approximation can be formalised using an asymptotic expansion in the aspect ratio of the filament $R/L\ll 1$. Here, we will proceed directly  by noting that since the rod is slender, bending it is easier than stretching or shearing it  (Love 1944). At the level of scaling, geometry implies that the out-of-line (bending) displacement of the centre line scales as $R$ while the axial displacement scales as $R^2/L$. 

At the surface of the filament no stress is applied.  Since the filament is thin this implies that
$\sigma_{yy} \approx \sigma_{zz} \approx 0$.  For a displacement field
${\bf u}=(u_x,u_y,u_z)$ equation (\ref{CE}) yields
\begin{eqnarray}
\sigma_{yy}=-\alpha p+(2\mu+\lambda)\partial_y u_y + \lambda(\partial_x u_x +\partial_z u_z)=0, \\
\sigma_{zz}=-\alpha p+(2\mu+\lambda)\partial_z u_z + \lambda(\partial_x u_x +\partial_y u_y)=0,
\end{eqnarray}
which can be solved for $\partial_yu_y$ and $\partial_zu_z$ to give
\begin{eqnarray}
\label{eliminate}
\partial_yu_y=\partial_zu_z=\frac{\alpha\, p-\lambda\partial_xu_x}{2(\mu+\lambda)}.
\end{eqnarray}
Equations (\ref{CE}) and (\ref{eliminate}) give the axial stress
\begin{equation}
\label{sf}
\sigma_{xx}=-\frac{\alpha \mu}{\lambda+\mu}p + \frac{3\lambda\mu+2\mu^2}{\lambda+\mu}\partial_xu_x.
\end{equation}
More specifically, when an infinitesimal axial element of the rod of length $dx$ is bent so that locally the
centreline curvature is $\partial_x \theta$,  fibres that are parallel to the neutral axis (coincident with the centre line for a homogeneous circular cross-section) and at a perpendicular distance $y$ from the neutral plane (defined by the neutral axis and the axis of bending) will be either extended or contracted by an amount $y\partial_x \theta dx$, so that the elastic strain $\partial_xu_x = -y\partial_x \theta$.  
This leads to an elastic stress that varies linearly across the cross-section; in addition there is a fluid pressure that is determined by (\ref{Cont}).  We insert
(\ref{eliminate}) into 
(\ref{Cont}) and find the evolution equation for the fluid pressure
\begin{equation}
\label{cont2}
k(\partial_{xx}p+\partial_{yy}p+\partial_{zz}p) = (\beta + \frac{\alpha^2}{\lambda+\mu})\partial_t p
-\frac{\alpha\mu}{\lambda+\mu}y\partial_{xt} \theta.
\end{equation}


To make the equations dimensionless we use the following definitions for the dimensionless primed variables:
\begin{eqnarray}
x = L\,x', ~~~~~~~ y = R\,y', ~~~~~~~ z = R\,z', ~~~~~~~ \theta = \frac{R}{L} \theta',  \nonumber \\
\sigma_{xx} = \frac{(2\mu^2 + 3\lambda\mu)R^2}{(\mu + \lambda)L^2}\,\sigma '_{xx}, ~~~
P = \frac{\pi(2\mu^2 + 3\lambda\mu)R^4}{4(\mu + \lambda)L^2}\,P', \nonumber \\
p = \frac{\alpha \mu R^2}{[(\mu+\lambda)\beta + \alpha^2]L^2}\,p', 
~~~~~ t = \frac{[(\mu+\lambda)\beta + \alpha^2]R^2}{(\mu+\lambda)k}\,t'.
\end{eqnarray}
and the immediately drop the primes, referring exclusively to dimensionless variables from now on.
We note that the axial stress, the compressive force and the pressure are scaled to reflect the dominance of bending deformations over all other modes, and the time is scaled to reflect the dominance of radial diffusion over axial diffusion.  

Then the stress in the filament, given by equation (\ref{sf}), can be written in dimensionless form as
\begin{eqnarray}
\label{S6}
\sigma_{xx} =  -y\,\partial_{x}\theta  -\frac{\delta}{4} p
\end{eqnarray}
Here, the first term reflects the purely elastic contribution well known from the theory of beams (Love; 1944), while the second term is proportional to the fluid pressure in the pores. The dimensionless parameter $ \delta = \frac{4\alpha^2\mu}{(2\mu + 3\lambda)[(\mu+\lambda)\beta + \alpha^2]} \sim O(1)$ for most materials denotes the ratio of the  fluid and solid stress. 

 \begin{figure}
\begin{center}
\includegraphics[width=7cm]{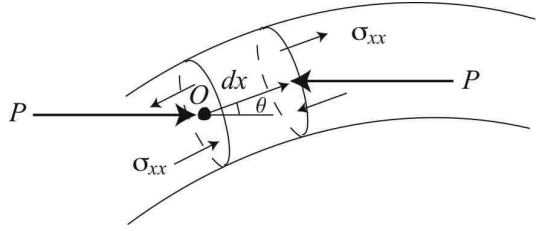}
\par
\caption{Schematic of the local torque balance in a bent rod under an externally applied compression $P$.  We balance torques about the point $O$, $x,y$ and $z$ are the coordinates in the body-fixed frame with $O$ as the origin.  Let $M(x)$ denote the total moment due to internal stresses generated in part by bending the elastic skeleton and in part by the fluid pressure field.  Balancing torques gives: $M(x+dl)-M(x)+{\bf dl}\times{\bf P} =0$.  In the limit $dl\to0$, $\partial_xM + P\sin\theta=0$. The total moment $M=E_pI\partial_x\theta+\alpha\int p\,y\,dA$, where $E_p$ is the effective elastic modulus of the poroelastic skeleton and $I$ is the moment of inertia of the cross section.}
\label{elastica}
\end{center}
\end{figure}

In the long wavelength approximation, it is preferable to use the stress resultant $F= \int \sigma_{xx} dA$ and the torque resultant $M=-\int y\sigma_{xx}  dA= \int y^2 \partial_x \theta dA+ \frac{\delta}{4}\int y p dA=M_e+M_f,$ as the variables of interest. Here $M_e$ is the elastic torque and $M_f$ is the fluid torque that arises due the transient effects of a pressure gradient across the filament. Then, local force and torque balances, which can be derived from (\ref{CE}), (\ref{F}) and (\ref{sf}), or equivalently directly (Figure \ref{elastica}), yield the dimensionless equations
\begin{eqnarray}
\label{Equil}
\partial_x F &=&0 \nonumber \\
\partial_x M + \frac{\pi}{4}P \sin \theta &=&0.
\end{eqnarray}

The first of these equations can be integrated immediately to yield $F =P$ with $P$ a constant determined by the boundary conditions. The second equation combines the effects of the elastic and fluid stresses that arise due to the fluid pressure, and requires the solution of the continuity equation (\ref{cont2}). For a rod with a circular cross-section, there is rotational symmetry in the problem. Choosing the  the axis of bending to coincide with the $z$ axis, 
we rewrite (\ref{cont2})
in polar coordinates $(r,\phi)$ using dimensionless variables as
\begin{equation}
\label{PP}
\partial_tp  - \frac{1}{r}\partial_r(r \partial_r p) - \frac{1}{r^2}\partial_{\phi\phi}p = r\,\sin\phi \,\partial_{xt}\theta.
\end{equation}
We see that the pressure in the fluid arises from the extensional and compressional stresses in the filament due to bending. 
The boundary conditions  for the pressure can be deduced using the following considerations: (a) the centre line of the rod does not suffer any deformation, and is symmetrically disposed, and (b) the pressure at the surface is determined by  the permeability of the surface layers and the flux through it. Then
\begin{eqnarray}
p = 0 ~~~ \hbox{at} ~~~ r=0, \nonumber \\
\label{bcp}
\hbox{Bi}\,p  + \partial_r p = 0 ~~~  \hbox{at} ~~~  r=1.
\end{eqnarray}
where Bi  = $\frac{\eta R}{k}$, and $\eta$ characterizes the flux through the surface for a given pressure drop (the ambient external pressure is assumed to be zero).   
The second boundary condition in (\ref{bcp}) on the pressure states that the flux of 
fluid through the surface is proportional to the pressure drop across the surface.
Bi = $\infty$  corresponds to a freely draining rod, where there is no pressure jump across the surface, and  Bi = 0 corresponds to a jacketed rod, which allows no flux through the surface. For a sponge, Bi$ >1$, while for  a plant (root, stem or leaf) Bi $< 1$ since it is designed to retain water. 

\begin{figure}
\begin{center}
\includegraphics[width=7cm]{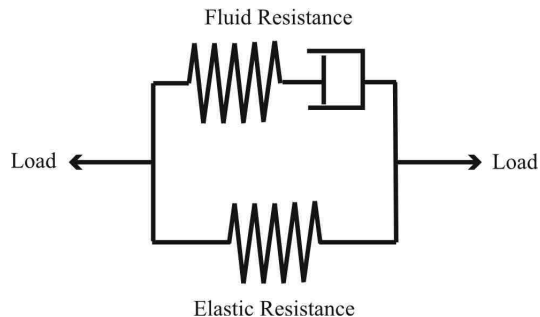}
\par
\caption{Mechanical analogue of the bending resistance of a poroelastic rod. For rapid displacements, the dashpot will not move and the fluid resistance due to the instantaneous pressure yields a response similar to a stiff (fluid) spring in parallel with an elastic spring. Eventually the dashpot will move to relieve the stress in the spring and the fluid resistance gradually
decays, leading to a purely elastic steady state.}
\label{mech}
\end{center}
\end{figure}

Expanding $p$ in terms of the homogeneous solutions of (\ref{PP}), we write
\begin{equation}
p=\sum_{m=0}^\infty\sum_{n=1}^\infty[A_{mn}\sin\,m\phi+B_{mn}\cos\,m\phi]J_m(r\sqrt{\lambda_{mn}})e^{-\lambda_{mn}t},
\end{equation}
where $A_{mn}$ and $B_{mn}$ are constants, $J_m$ is the Bessel function of order $m$, and $\lambda_{mn}$ is determined by the boundary conditions.
Inspection of the inhomogeneous term on the RHS of equation (\ref{PP}) yields $m=1$ so that $A_{1n}=A_n$, $B_{mn}=0$, and $\lambda_{1n}=\lambda_n$. Since the boundary condition is a linear combination of $p$ and $\partial_r p$  we are guaranteed to have a complete basis. We therefore look for a solution to the inhomogeneous equation  (\ref{PP}) of the form
\begin{equation}
\label{ep}
p=\sum_{n=1}^\infty A_n(t)\sin\,\phi J_1(r\sqrt{\lambda_n}),
\end{equation}
where $\lambda_n$ is determined by substituting (\ref{ep}) into (\ref{bcp}) which yields 
\begin{equation}
\label{lam}
\partial_rJ_1(r\sqrt{\lambda_n}) + \hbox{Bi}\,J_1(r\sqrt{\lambda_n}) = 0 ~~~~ \hbox{at} ~ r=1.
\end{equation}
%
Inserting (\ref{ep}) into (\ref{PP}) yields
\begin{equation}
\label{13}
\sum_{n=1}^\infty (\partial_tA_n+\lambda_nA_n)J_1(r\sqrt{\lambda_n}) = r\partial_{xt}\theta.
\end{equation}
Multiplying (\ref{13}) by $r\,J_1(r\sqrt{\lambda_{n'}})$ and integrating across the cross-section
gives
\begin{equation}
\label{A_n}
\partial_tA_n + \lambda_nA_n =  \chi_n \partial_{xt} \theta,
\end{equation}
where
\begin{equation}
\chi_n = \frac{\int_0^1r^2J_1(r\sqrt{\lambda_n}) dr}{\int_0^1r [J_1(r\sqrt{\lambda_n})]^2 dr}.
\end{equation}
Solving equation (\ref{A_n}) yields
\begin{equation}
A_n=\chi_n\int_0^te^{-\lambda_n(t-t')}\partial_{xt'}\theta dt',
\end{equation}
so that  (\ref{ep}) may be rewritten as:
\begin{equation}
p = \sum_{n=1}^\infty  \chi_n \sin\phi \,J_1(r\sqrt{\lambda_n})\int_0^te^{-\lambda_n(t-t')}\partial_{xt'}\theta dt'.
\end{equation}
Then (\ref{S6}) allows us to write the total axial stress $\sigma_{xx}$ at a cross-section as 
\begin{equation}
\sigma_{xx} = -r\sin\phi\, \partial_x\theta -\frac{\delta}{4}\sum_{n=1}^\infty  \chi_n \sin\,\phi J_1(r\sqrt{\lambda_n})\int_0^te^{-\lambda_n(t-t')}\partial_{xt'}\theta dt'.
\end{equation}
The dimensionless torque resultant is given by 
\begin{eqnarray}
\label{fmom}
M =  - \int r\sin\phi \,\sigma_{xx}\,dA
=\frac{\pi}{4}\partial_x\theta+\frac{\pi\delta}{4} \sum_{n=1}^\infty \gamma_n \int_0^te^{-\lambda_n(t-t')}\partial_{xt'}\theta dt',
\end{eqnarray}
where $\gamma_n = \chi_n \int_0^1r^2J_1(r\sqrt{\lambda_n}) dr$.
Substituting the result into the equation for torque balance (\ref{Equil}) yields the dimensionless equation for the poroelastica (see Figure \ref{elastica})
\begin{equation}
\label{poro}
\partial_{xx}\theta + P\, \sin\theta + \delta \sum_n^\infty \gamma_n \int_0^te^{-\lambda_n(t-t')}\partial_{xxt'}\theta dt'=0,
\end{equation}
The first two terms correspond to the usual terms in the classical elastica (Love 1944)  for the bending of a rod with a circular cross section, while the final term is due to the instantaneous fluid pressure not being equilibrated across the cross section.  The influence of the fluid is to create a material with ``memory'', so that the current state of the filament is determined by its entire history.  The kernel in the memory function for the fluid resistance is $e^{-\lambda_n(t-t')}$ so that the fluid resistance is analogous to that of a Maxwell fluid (Bird, Armstrong and Hassager 1987) with relaxation times $1/\lambda_n$ which measure the rate of decay of the $n^{th}$ transverse mode in response to the rate of change of the curvature of the filament $\partial_{xt} \theta$. A mechanical analogue of the resistance of a poroelastic filament is presented in Figure \ref{mech} and shows the connection to simple viscoelastic models.   
 
The dynamics of a poroelastic rod are then determined by the solution of the equation for local torque balance (\ref{poro}), subject to the boundary condition on the fluid pressure at the surface (\ref{lam}) which determines the decay constants $\lambda_n$, and additional boundary conditions on the ends of the rod, which we now consider in some specific cases.

\section{Planar load controlled buckling}
\label{Pcontrolled}

When an initially straight  rod that is simply supported at either end is subject to a constant compressive force  $P$ applied suddenly at $t=0$, the boundary conditions at the ends are   
\begin{eqnarray}
\partial_x\theta(0,t)&=&0, \nonumber \\ 
\partial_x\theta(1,t)&=&0, \label{BC}
\end{eqnarray}
and the initial  condition is
\begin{equation}
\label{ic}
\theta(x,0) = 0.
\end{equation}

The complete time evolution of the rod is then given by the solution of the integrodifferential equation (\ref{poro}) subject to the boundary conditions (\ref{BC}), the initial conditions (\ref{ic}) and the condition (\ref{lam}) which determines the rate constants $\lambda_n$.

\subsection{Short time behaviour, $t\ll1$}
\label{short}

Expanding the solution about the initially straight state $\theta=0$, we write 
\begin{equation}
\theta = \epsilon \theta_1(t) \label{stexp}
\end{equation}
where $\epsilon \ll1$. Substituting the expression (\ref{stexp}) into equation (\ref{poro})  and linearizing yields
\begin{equation}
\label{lin}
\partial_{xx}\theta_1 + P\,\theta_1 + \delta \sum_{n=1}^\infty \gamma_n\int_0^te^{-\lambda_n(t-t')}\partial_{xxt'}\theta_1 dt'=0,
\end{equation}
subject to the boundary conditions
\begin{equation}
\partial_x\theta_1(0,t)=\partial_x\theta_1(1,t)=0. \label{ibc}
\end{equation}

\begin{figure}
\begin{center}
\includegraphics[width=8cm]{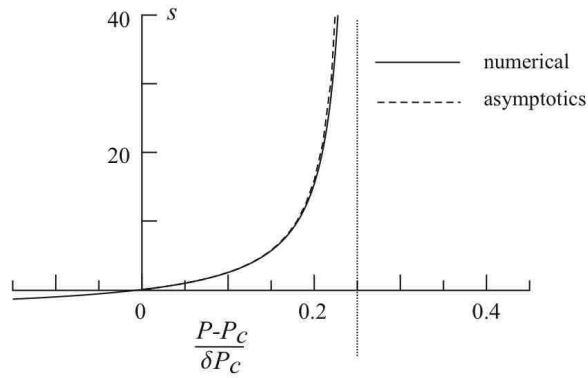}
\par
\caption{Growth rate $s$ of the deflection as a function of the dimensionless external load $P$; $P_c=\pi^2$ is the critical compression above which a simply supported purely elastic rod buckles. Here the surface permeability parameter Bi$=0.1$. When $\frac{P-P_c}{\delta P_c}=\frac{1}{4}$ the growth rate becomes infinite and one must consider inertial effects, which are neglected here. For comparison, we show the numerical results obtained by solving (\ref{poro}) with initial conditions $\theta(0)=0.001\,\cos\,\pi x$,  $\theta(dt)=0.001\,e^{s\,dt} \cos\,\pi x$, where $s$ is the theoretically calculated exponent, with $dt=0.001$, $dx=0.01$. 
}
\label{s(p)}
\end{center}
\end{figure}

To solve (\ref{lin}-\ref{ibc}) we use separation of variables writing $\theta_1(x,t) = g(x)f(t)$ and substituting the result into (\ref{lin}) to obtain two equations for $g(x)$ and $f(t)$.   The function $g(x)$ is determined by the solution of the eigenvalue problem :
\begin{equation}
\label{21}
(1+\xi)\partial_{xx}g + P\,g=0, ~~~ \partial_xg(0)=0, ~~~ \partial_xg(1)=0.
\end{equation}
Here the separation constant $\xi=(P-\pi^2)/\pi^2$ is the relative difference between the applied load $P$ and the dimensionless  buckling load, $P_c = \pi^2$ for a purely elastic rod that is simply supported at its ends.  The function $f(t)$ satisfies  
\begin{equation}
\label{22}
\frac{\xi f(t)}{\delta}=\sum_{n=1}^\infty \gamma_n\int_0^te^{-\lambda_n(t-t')}\partial_{t'}f (t')dt',
\end{equation}
Using Laplace transforms (${\cal L} (f(t))=\int_0^\infty e^{-st}f(t)\,dt$) we solve (\ref{22}) and find that
\begin{equation}
\label{res}
f(t)=-\sum_{s\in S}e^{st}f(0)\sum_{n=1}^\infty\frac{\gamma_n}{\lambda_n+s}.
\end{equation}
where $f(0)$ is determined by the initial condition and the set $S$ is composed of 
elements which satisfy  
\begin{equation}
\label{sss}
\frac{\xi}{\delta}-\sum_{n=1}^\infty\frac{\gamma_ns}{\lambda_n+s}=0.
\end{equation}
The growth rate at the onset of the dynamic buckling instability is therefore given by the largest $s$ that satisfies (\ref{sss}). In Figure \ref{s(p)}, we plot the growth rate $s$ as a function of the rescaled external load $\frac{\xi}{\delta} =\frac{P-P_c}{\delta P_c}$, with $P_c = \pi^2$, obtained by solving (\ref{21}) and (\ref{sss}).   When $s<0$ we do not have an instability, corresponding to the case when $P<P_c$. In the poroelastic regime, when $P_c<P<P_c(1+\frac{\delta}{4})$,  fluid flows across the filament in response to the stress gradient in the transverse direction,  and the phenomenon is qualitatively different from the buckling of a purely elastic rod.
Since the time it takes fluid to flow across the filament is longer than it takes a bending wave to propagate 
the length of the filament, poroelastic 
buckling is sometimes called creep buckling (Biot 1964).

We now turn to the dependence of the buckling transition on the surface permeability parameter Bi.  Substituting a pressure field of the form $p(x,r,\phi,t)=h(r)\partial_x\Theta(x)\sin\phi\,e^{st}$ 
and $\theta = e^{s\,t}\Theta(x)$
into (\ref{PP}) yields
\begin{equation}
\label{PD2}
s\,h -\frac{1}{r}\partial_r(r\partial_rh) + \frac{h}{r^2} = s\,r,
\end{equation}
subject to boundary conditions (\ref{bcp}) which are now given by
\begin{eqnarray}
h= 0 ~~~ \hbox{at} ~~~ r=0, \nonumber \\
\label{bcp2}
\hbox{Bi}\,h  + \partial_r h = 0 ~~~  \hbox{at} ~~~  r=1,
\end{eqnarray}
Thus (\ref{PD2}) yields $\lim_{s\to0}h = 0$ and $\lim_{s\to\infty}h=-r$ corresponding to the case of infinitely slow and infinitely fast growth rates respectively. Consequently, for infinitely slow buckling the fluid supports no load.  In the case $s\to\infty$ near $r=1$ a boundary layer emerges where the internal solution ($h=r$) is matched to the boundary condition (\ref{bcp2}) at $r=1$.  Balancing the first two terms of (\ref{PD2}) the length scale of the boundary layer $l_{bl}\sim1/\sqrt s$ or in dimensional terms $l_{bl}\sim \sqrt{\frac{(\mu+\lambda)k}{s[(\mu+\lambda)\beta+\alpha^2]}}$.

\begin{figure}
\begin{center}
\includegraphics[width=8cm]{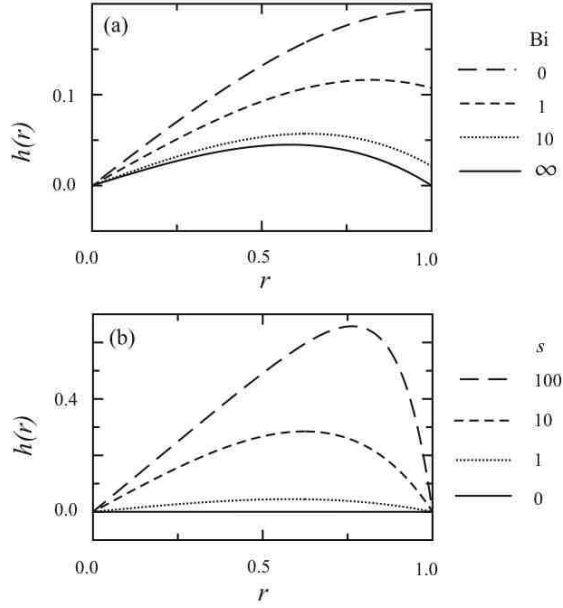}
\par
\caption{The radial variation of the fluid pressure at the onset of buckling, $h(r)$,  (a) for growth rate $s$ = 1 and various 
values of the surface permeability parameter Bi (larger Bi corresponds to a more permeable surface) and (b) for Bi = $\infty$ and various $s$.}
\label{pchbi}
\end{center}
\end{figure}

To complement these asymptotic results,  we solve  (\ref{PD2}-\ref{bcp2}) numerically and plot the radial variation in the pressure. In Figure \ref{pchbi}a we show $h(r)$ for $s=1$ and various Bi and in Figure \ref{pchbi}b we show $h(r)$ for Bi = $\infty$ and various $s$.  As expected, we see that as the surface permeability increases  ({\it i.e.} Bi increases) for a given growth rate of the instability (corresponding to a given load) the pressure variations across the filament decrease. 
On the other hand, as the growth rate increases, a boundary layer appears in the vicinity of the free surface of the rod to accommodate the slow permeation of fluid in response to the stress gradients.


Having considered  the onset of poroelastic buckling we now turn to the transition from poroelastic to inertial dynamics which occurs for very large compressive loads when the fluid cannot move rapidly enough to keep up with the elastic deformations. 
%
%
For large $s$ the condition that determines the growth rate (\ref{sss}) reads
\begin{equation}
\label{sim}
0 \approx \frac{\xi}{\delta}-\sum_n\gamma_n(1-\lambda_n/s).
\end{equation}
%
We have computed $\sum_n \gamma_n=1/4$ for integrals of Bessel functions.
Using the definition of the separation constant $\xi=P/P_c-1$  we solve equation (\ref{sim}) for the growth rate:
\begin{equation}
s \approx \frac{\sum_n\gamma_n\lambda_n}{\frac{P-P_c}{\delta P_c} - \frac{1}{4}},
\end{equation}
showing that it indeed diverges when $\frac{P-P_c}{\delta P_c} \to \frac{1}{4} $ consistent with Figure \ref{s(p)}.  


\subsection{Long time dynamics $t\gg1$}
\label{long}

\begin{figure}
\begin{center}
\includegraphics[width=7cm]{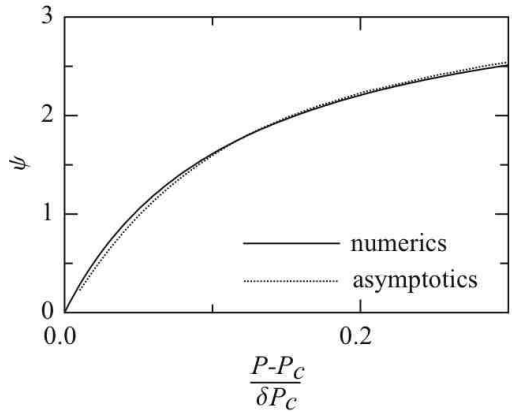}
\par
\caption{As the solution approaches the equilibrium shape the difference between the current and equilibrium shape decays exponentially with a rate $\psi$, which is plotted against the applied compression. $\delta=1$, Bi = 0.1, $dx=0.01$, 
$dt=0.01$.  The numerical computation is begun with $\theta = 0.9\,\theta_0$.
}
\label{exp}
\end{center}
\end{figure}

In the long time limit $t\gg1$, i.e. when the fluid has enough time to diffuse across and along the filament, the shape of the filament approaches that of the ideal {\it elastica}. To capture the dynamics of this process, we linearise (\ref{poro}) about the steady state solution by letting 
\begin{equation}
\theta = \theta_0(x) + \epsilon \theta_1(x,t) \label{ltexp}
\end{equation}
with $\epsilon \ll 1$. Substituting the expansion (\ref{ltexp}) into (\ref{poro}), at leading order we get 
\begin{equation}
\partial_{xx}\theta_0 + P\,\sin\theta_0=0.
\end{equation}
At $O(\epsilon$), we get
\begin{equation}
\label{lt}
\partial_{xx}\theta_1 + P\,\cos(\theta_0)\,\theta_1 + \delta\sum_n\gamma_n\int^te^{-\lambda_n(t-t')}\partial_{xxt'}\theta_1dt' =0.
\end{equation}
To simplify the equations further we consider the convolution integral in (\ref{lt}) for typical values of Bi =  0.1, corresponding to the case for soft gels and biological materials. Then (\ref{lam}) yields \{$\lambda_n$\} = \{3.67, 28.6, 73.1, 137, 221, 325, ...\} and $\sum_{n=2}^\infty \gamma_n/\gamma_1 = 0.0152$. Given the large separation between the decay constants, we see that the dominant contribution in the integral arises from $\lambda_1$ leading to an approximation of
(\ref{lt}) that reads
\begin{equation}
\partial_{xx}\theta_1 + P\,\cos(\theta_0)\,\theta_1 + \delta\gamma_1\int^te^{-\lambda_1(t-t')}\partial_{xxt'}\theta_1dt'=0.
\end{equation}
Using separation of variables, $\theta_1(x,t) = g(x)f(t)$, we find that $g(x)$ is given by the solution of the eigenvalue problem
\begin{equation}
(1-\xi)\partial_{xx}g + P\,\cos(\theta_0)\,g = 0, ~~~ \partial_xg(0)=0, ~~~ \partial_xg(1)=0,
\end{equation}
while the temporal part $f(t)$ satisfies 
\begin{equation}
\label{lt2}
-f\xi = \delta\gamma_1\int^te^{-\lambda_1(t-t')}\partial_{t'}f\,dt'.
\end{equation}
Since we are interested in the asymptotic behaviour for $t\gg1$ we multiply 
both sides of equation (\ref{lt2}) by $e^{\lambda_1t}$ and differentiate with respect to time to find 
\begin{equation}
\label{psi}
\partial_t f = \frac{-\lambda_1\xi}{\delta\gamma_1+\xi}f ~ \equiv -\psi f 
\end{equation}
We observe that the poroelastic solution approaches the elastic steady shape exponentially fast at late times. In Figure \ref{exp}, we plot the exponent $\psi=\lambda_1\xi /(\delta\gamma_1+\xi)$ versus $(P-P_c)/\delta P_c$ and see that the larger the value of $P$ the faster the solution approaches the final shape.

\subsection{Intermediate time dynamics}
\label{numerics}

\begin{figure}
\begin{center}
\includegraphics[width=8cm]{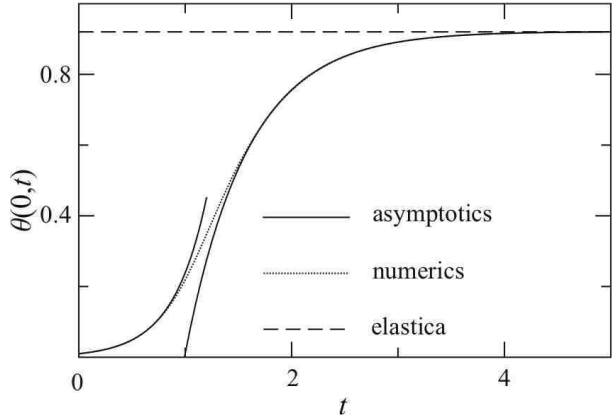}
\par
\caption{$\theta(0,t)$ for $P=11$, $\delta=1$, Bi=0.1, $dx=0.01$, $dt=0.001$ and $\theta(x,0)=\theta(x,dt)=0.01\,\cos\,\pi x$. 
The short time asymptotic is for a growth rate $s(P)$ found from equation (\ref{sss}).  The long time asymptotic is of the form $Ae^{-\psi \, t} + \theta_0$, where
$\psi$ is the rate of decay to the equilibrium angle $\theta_0$, and $A$ is a fitting parameter.}
\label{full}
\end{center}
\end{figure}

\begin{figure}
\begin{center}
\includegraphics[width=10cm]{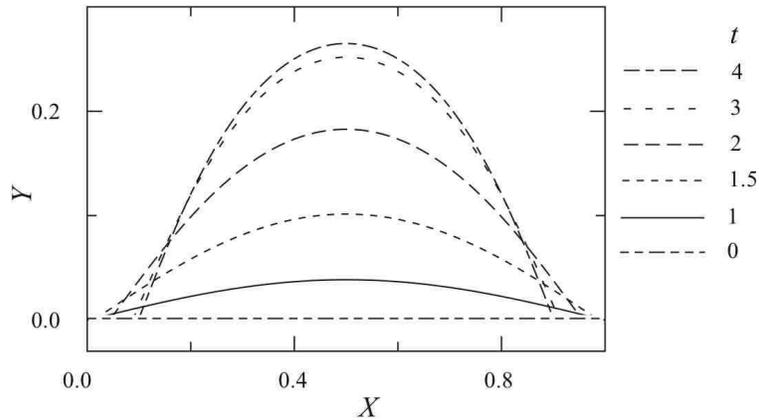}
\par
\caption{Shape of the buckling filament $X(t)$, $Y(t)$ for $P=11$,
in the laboratory frame as a function of time for $P=11$, $\delta=1$, Bi=0.1, $dx=0.01$, $dt=0.001$ and $\theta(t=0)=\theta(t=dt)=0.01\,\cos\,\pi x$. }
\label{bs}
\end{center}
\end{figure}

For intermediate times we have to solve for the shape of the poroelastica numerically.  Our arguments in the previous section allow us to neglect the contributions from the higher modes so that a good approximation to (\ref{poro}) is given by
\begin{equation}
\partial_{xx}\theta + P \sin\,\theta + \gamma_1\int_0^te^{-\lambda_1(t-t')}\partial_{xxt'}\theta dt' = 0. \label{pesim}
\end{equation}
For ease of solution, we convert the integrodifferential equation to a partial differential equation
by multiplying  (\ref{pesim}) by $e^{\lambda_1t}$ and differentiating with respect to time, so that
\begin{equation}
\label{solve}
(1+\gamma_1)\partial_{xxt}\theta + \lambda_1\partial_{xx}\theta + P\,\cos\,\theta\,\partial_t\theta + \lambda_1\,P\,\sin\,\theta = 0.
\end{equation}
We solve equation (\ref{solve}) subject to the boundary conditions (\ref{BC}) using a Crank-Nicolson finite difference scheme in space and we extrapolate the non-linearity using the previous two time steps. This gives us a scheme with second order accuracy in time.  For a time step $dx=0.01$ and a space step $dt=0.001$ the difference between the 
numerical and analytical initial growth rate is 0.2\% (see Figure \ref{s(p)}).
In Figure \ref{full} we show the variation of the angle $\theta(0,t)$ determined using the numerical simulation for the case when the dimensionless buckling load is slightly larger than the threshold for the poroelastic buckling, with $(P-P_c)/P_c \sim 0.17$. For comparison, we  also show the asymptotic solutions for short and long times determined in the previous sections, and find that they agree well with the numerical solution.  
To determine the shape of the filament we use the kinematic relations 
\begin{equation}
\partial_xX=\cos\theta, ~~~\partial_xY=\sin\theta,
\end{equation}
where $X(x,t)$ and $Y(x,t)$ are the position of the centreline.
Figure \ref{bs} shows the shape of the filament as it evolves from the initially unstable straight shape to the final elastic equilibrium via a transient overdamped route. In sharp contrast, a purely elastic rod subject to the same initial and boundary conditions would  vibrate about the final state forever (in the absence of any damping).  

\section{Displacement controlled planar buckling}
\label{dcb}

\begin{figure}[h]
\begin{center}
\includegraphics[width=8cm]{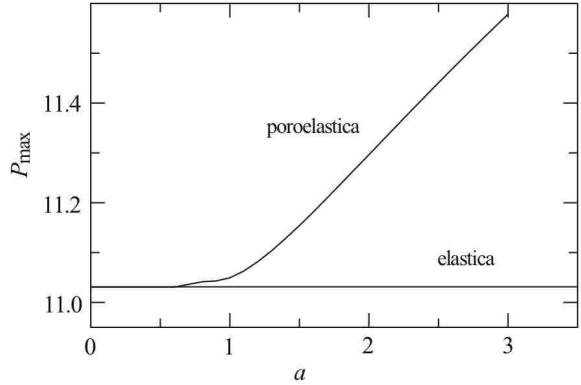}
\par
\caption{Maximum compressive load , $P_{max}$, for displacement controlled buckling, where the displacement field is given by 
$\Delta(t) \approx 0.1[1+\tanh\,at]$. Bi = 0.1, $\delta=1$, $dt=0.01$, $dx=0.01$.}
\label{Pmax}
\end{center}
\end{figure}

In many problems involving instabilities, there is a qualitative difference between load controlled and displacement experiments.
To understand the difference we consider the problem of displacement controlled buckling of a poroelastic filament and compare the results with those of the previous section. Since the centre line of the filament is assumed to be inextensible,  the change in the end-to-end distance is given by
\begin{equation}
\label{constraint}
\Delta(t) = 1 - \int_0^1\cos\,\theta(x,t)\,dx.
\end{equation}
We choose the functional form  
\begin{equation}
\label{Delta1}
\Delta(t)=\frac{\Delta_{max}}{2}[1+\tanh\,at],
\end{equation}
to allow us to ramp up the displacement to a maximum amplitude $\Delta_{max}$ 
at a characteristic rate $\frac{1}{2}\Delta_{max} a$.
The shape of the poroelastica is now determined by the solution of (\ref{solve}), (\ref{constraint}) and
(\ref{Delta1});  the unknown load $P(t)$ is now determined at every time step by using an iteration method to enforce (\ref{constraint}). For an initial guess to start this procedure, we note that after the onset of buckling when $P>P_c$, for small amplitudes  $\theta = \epsilon\,\cos\pi x$ ($\epsilon\ll1$) is a solution
of (\ref{solve}) and (\ref{BC}).  Substituting into (\ref{constraint}) gives
\begin{equation}
\label{Delta}
\Delta = 1 - \int_0^1\cos(\epsilon\,\cos\pi x)\,dx \approx \frac{\epsilon^2}{4}.
\end{equation}
\begin{figure}
\begin{center}
\includegraphics[width=8cm]{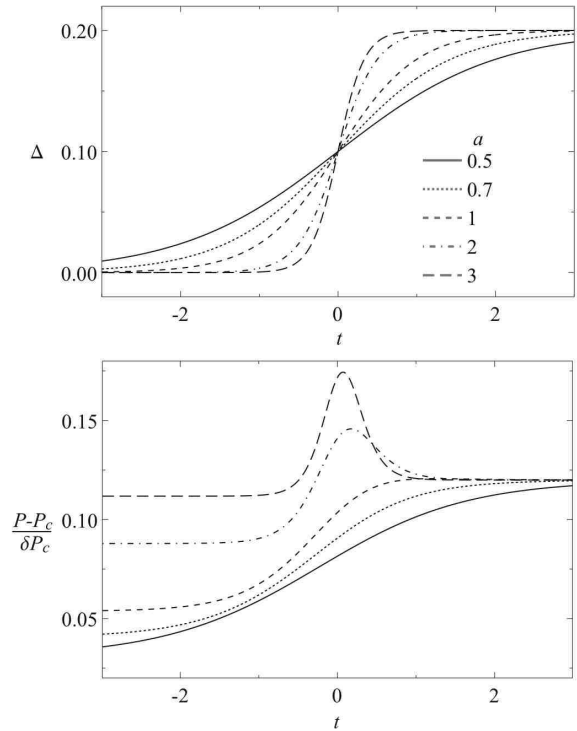}
\par
\caption{$\frac{P-P_c}{\delta P_c}$, where $P$ is the load and $P_c$ is the critical load required
for buckling, corresponding to a change in end to end displacement $\Delta(t)\approx 0.1[1+\tanh\,at]$, for various $a$.  
For some later times the more quickly applied displacement
corresponding to larger $a$ requires a lower compressive force. The graphs correspond to the following parameter values:
$dx=0.01$, $dt=0.002$, Bi=0.1, $\delta=1$, $\theta(-3)= \theta(-3+dt)=2\sqrt{\Delta(-3)}\,\cos\,\pi x$.}
\label{dispa}
\end{center}
\end{figure}
Therefore, we choose $\theta(x,t_0) \sim 2\sqrt{\Delta(t_0)}\,\cos\,\pi x$.  In Figure \ref{dispa}b, we show the evolution of the load $P(t)$ for various values of $a$. 
$P$ is roughly constant for very short and very long times, but  changes as $\Delta$ varies quickly for intermediate times.   
We can understand the initial plateau by considering the case when ($e^{at}\ll1$), so that  (\ref{Delta1}) yields
\begin{equation}
\Delta = \frac{\Delta_{max}}{2}[1- \frac{1-e^{2at}}{1+e^{2at}}] \approx \Delta_{max}e^{2at}.
\end{equation}
In light of the geometrical constraint (\ref{Delta}) valid for small displacements, this yields 
\begin{equation}
\theta \approx 2\sqrt{\Delta_{max}}e^{at}\,\cos\,\pi x.
\end{equation}
Comparing this with the short time behaviour of a poroelastic filament considered in \S 4\ref{short} we see that exponential growth of small angles corresponds to a constant compressive force seen in Figure \ref{dispa}. A similar argument holds for late times, when the system relaxes to its purely elastic equilibrium. For intermediate times, the load can be larger than that for the case of a purely elastic filament. 
The difference in the loads is due to the fluid resistance in the porelastica.  A way of visualising this is shown in Figure \ref{Pmax}. For slowly applied displacement fields $P_{max}$  is almost the same for the elastic and poroelastic cases; however, for rapidly applied displacements, corresponding to large values of $a$, the compressive force in the poroelastic case is larger due to fluid resistance arising from the pressure gradients across the bending filament.  

\section{Filament embedded in an external medium and subject to axial torque and axial thrust}
\label{twist}

\begin{figure}
\begin{center}
\includegraphics[width=10cm]{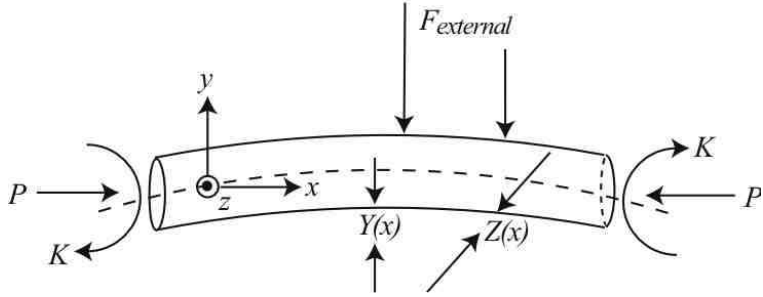}
\par
\caption{Schematic diagram of a rod buckling under an applied twist and compression in an external medium.  $Y(x,t)$ and $Z(x,t)$ are the displacements of the centre line in the $y$ and $z$ directions respectively. The dotted line denotes the axis of symmetry.}
\label{schem2}
\end{center}
\end{figure}

We finally turn to the case of a rod embedded in an external medium subject to an axial moment, $K$, and a compressive force $P$ (see Figure \ref{schem2}).  The presence of the twist  causes the instability to become non-planar, and the filament adopts a helical conformation; the presence of an external medium typically causes the instability to manifest itself with a higher wave number than otherwise.  Letting  the displacements of the centre line  in the $y$ and $z$ directions be $Y(x,t)$, $Z(x,t)$  respectively, we scale the  kinematic variables accordingly to define the dimensionless displacements   
\begin{equation}
Y=R\,Y',~~~~Z=R\,Z'.
\end{equation}
The dimensionless axial moment is defined as $K = \frac{\pi(2\mu^2 + 3\lambda\mu)R^4}{4(\mu + \lambda)L}K'$.  
If the transverse displacements of the filament are small, the resistance of the external medium can be well approximated using the response of a linear Hookean solid. In light of  the analogy between linear elasticity and Stokes flow, we can use the results of classical slender body theory in hydrodynamics (Batchelor 1970, Cox 1970) and write the vector of dimensionless external forces on the filament as 
${\bf F'}_{external}=(0,-\pi E\,Y/4,-\pi E\,Z/4)$, where the dimensionless parameter $E$  is given by
\begin{equation}
E = \frac{16\mu_mL^2(\mu + \lambda)}{\hbox{ln}(\frac{L}{R})\,R^2(2\mu^2 + 3\mu\lambda)},
\end{equation}
where $\mu_m$ is the Lam\'e coefficient of the surrounding medium. This approximation is valid when $R/L \ll 1$, a condition consistent with the geometry of a thin filament.    We will further assume that the filament is free to rotate in the medium, i.e. there is no torque resisting this mode of motion, which varies in any case as $R^2$ and is thus negligible in most situations.


To derive the evolution equation for the shape of the filament we use the constitutive equation (\ref{CE}) to write down equations for the balance of forces in the $y-$ and $z-$directions as for the
planar filament. After dropping primes this  leads to
\begin{eqnarray}
\label{N1}
\partial_{xxxx}Y + K\partial_{xxx}Z +  \delta\sum_{n=1}^\infty \gamma_n \int_0^te^{-\lambda_n(t-t')}\partial_{xxxxt}Y dt' + P\partial_{xx}Y + E\,Y=0, \\
\label{N2}
\partial_{xxxx}Z - K\partial_{xxx}Y +  \delta\sum_{n=1}^\infty \gamma_n \int_0^te^{-\lambda_n(t-t')}\partial_{xxxxt}Z dt' + P\partial_{xx}Z + E\,Z =0, 
\end{eqnarray}
where, since equation (\ref{PP}) is linear,
 we have superposed the two solutions for bending in the $y-$ and $z-$directions.
Taking $\zeta = Y + i\,Z$ equations (\ref{N1}) and (\ref{N2}) may be written as a single equation for the complex variable $\zeta$
\begin{equation}
\label{N3}
0=\partial_{xxxx}\zeta - i\,K\partial_{xxx}\zeta + P\partial_{xx}\zeta + E\,\zeta  + \delta \sum_{n=1}^\infty \gamma_n \int_0^te^{-\lambda_n(t-t')}\partial_{xxxxt'}\zeta dt'.
\end{equation}
For a simply supported filament, the four boundary conditions are 
\begin{eqnarray}
\label{boun}
 \zeta(0) = \zeta(1) = \partial_{xx}\zeta(0)-i\,K\partial_x\zeta(0)=\partial_{xx}\zeta(1)-i\,K\partial_x\zeta(1)  = 0.
\end{eqnarray}
We can treat equations (\ref{N3})-(\ref{boun}) in exactly the same fashion as  the planar problem and use separation of variables $\zeta(x,t)=g(x)f(t)$ to get
\begin{eqnarray}
\label{38}
(1+\xi)\partial_{xxxx}g - i\,K\partial_{xxx}g + P\partial_{xx}g + E\,g =0,\nonumber \\
g(0)=g(1)=\partial_{xx}g(0)-i\,K\partial_xg(0)=\partial_{xx}g(1)-i\,K\partial_xg(1)=0,
\end{eqnarray}
an eigenvalue problem for the separation constant $\xi$ and $g(x)$. The temporal part of the solution $f(t)$ satisfies
\begin{equation}
\frac{\xi}{\delta}-\sum_{n=1}^\infty\frac{\gamma_ns}{\lambda_n+s}=0.
\end{equation}
which is the same as equation (\ref{lt2}) for the temporal part of the solution for planar buckling. Thus, once the separation constant $\xi$ is found,  equation (\ref{sss}) yields the growth rate $s(P,K,E,\delta,\hbox{Bi})$  as a function of the loading parameters and the material constants (see Figure \ref{s(p)}). 

As an example of how the influence of an external medium can lead to higher modes becoming unstable at lower compressions than the fundamental mode we consider equation (\ref{38}) in the case $K=0$
\begin{equation}
\label{sep}
(1+\xi)\partial_{xxxx}\zeta  + P\partial_{xx}\zeta + E\,\zeta =0,
\end{equation}
which is an eigenvalue problem for $\xi$ and $\zeta$ for a given $P$.  At the ends ($x=0$ and $x=1$) the displacements and bending moments vanish so that the boundary conditions associated with (\ref{sep}) are:
$\zeta(0)=\partial_{xx}\zeta(0)=\zeta(1)=\partial_{xx}\zeta(1)=0$.
The only nonzero solutions to (\ref{sep}) occur when $\zeta = \sin\,q_nx$, where $q_n = n\pi,$ $n=1,2,...$   The critical compression $P_c(n)$ where the $n^{th}$ mode becomes unstable is found to be (Landau \& Lifshitz 1970)
\begin{equation}
P_c(n) = \pi^2n^2 + \frac{\hbox{E}}{\pi^2n^2}.
\end{equation}
We see that the critical buckling load for a given mode number $n$ increases as the stiffness of the environment $E$ increases. Furthermore, for large $E$, the chosen mode shape does not correspond to the fundamental mode $n=1$, since $\partial P_c/\partial n =0$ yields $n=\frac{1}{\pi}E^{1/4}$ for an infinite rod. Physically this occurs because short wavelength modes do not deform the stiff elastic environment as much, while the penalty associated with a higher curvature is not too much of a price to pay.

\begin{figure}
\begin{center}
\includegraphics[width=7cm]{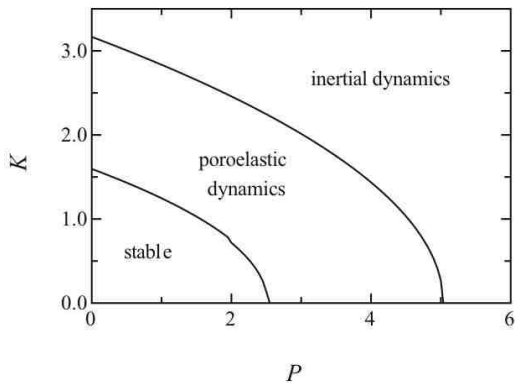}
\par
\caption{For the case of an applied compression $P$ and axial twisting moment $K$, we show the three short time regimes (stable, poroelastic, and inertial) as in Figure \ref{s(p)}. $E=\delta=1$.}
\label{SPK}
\end{center}
\end{figure}

For the case when $K \ne 0$, we cannot solve the eigenvalue problem analytically, and present the results using a phase diagram shown in Figure \ref{SPK}. We find three distinct regimes: for $s<0$ the system is stable; for $0<s<\infty$ we have the poroelastic regime where the system buckles on the same time scale as the fluid pressure diffuses; finally, we have the elastic regime where the system buckles so fast that the fluid does not move and all the deformation occurs in the solid skeleton.  

\section{Discussion}
\label{disc}

The usefulness of poroelastic theory is limited to a range of time scales.  
The poroelastic time scale associated with decay of pressure fields is $\tau_{p} \sim \frac{\alpha^2 R^2}{\mu\, k}$, recalling that  $\alpha$ is the 
fluid volume fraction, $R$ is the smallest macroscopic length scale of the system ($R\gg l_p$), $\mu$ is the effective Lam\'e coefficient of the composite material,
and $k \sim \frac{l_p^2}{\rho\nu}$ is the matrix permeability.   
If the time scale of the forcing, $\tau \ll \tau_{p}$ the fluid will not move relative to the solid and Hookean elasticity and the effects of inertia are sufficient to describe the system adequately.  If $\tau \gg \tau_{p}$, the fluid pressure will equilibrate with the surroundings and once again classical elasticity suffices to describe the system, albeit with different Lam\'e coefficients. 
However, if $\tau \sim \tau_{p}$ the dynamics will be governed by poroelasticity.

Biological systems are composed mainly of fluid, so poroelasticity will be applicable at some time scale (see Table 2 for estimates).  Furthermore, they are
characterized by extreme geometries (e.g. beams, plates and shells), which  
led us to consider in detail the dynamics of slender poroelastic objects, and particularly the buckling of a planar filament. 
Biological materials are usually anisotropic and we expect the permeability and elasticity 
tensors to reflect this feature.  Taking $k_l$ to be the 
permeability in the axial direction, we can neglect axial diffusion if $\frac{k_l R^2}{k L^2}\ll 1$.
The opposite limit, where $\frac{k_l R^2}{k L^2}\gg 1$, has been studied by Cederbaum et al. (2000).
The dynamical behavior of these objects is separated into two different regimes, 
one governed by fast inertial effects, and the other by the slow dynamics of fluid flow. 
These regimes are of course well known in bulk materials (Biot 1957), but here they appear in a slightly different guise due to the effect of the slender geometry of the system.  
The onset of planar poroelastic load controlled buckling was first considered by Biot (1964).  In this paper we broaden and deepen the understanding of this
phenomenon.  
An important outcome of our studies is the {\it poroelastica} equation, which is a simple integro-differential equation with one time constant that 
describes the dynamics of a poroelastic filament under a compressive load. 
The bending resistance of the filament is analogous to a (fictional) Maxwell material, where the time constant 
is the rate at which the pressure field decays (determined by the material parameters and the geometry). 
We then used this equation to study not only the
onset of buckling, but also the entire dynamics  up until saturation for both load controlled and displacement controlled buckling.
%

A series of three-point-bending experiments (Scherer 1992; Scherer 1996) have shown that the 
mechanical response of a silica gel rod immersed in acetone or ethanol can be described using
poroelastic theory.  The theory developed by Scherer (1992) applies only to situations where
the displacement is applied much faster than the poroelastic time scale and is a special case of 
the more general 
theory presented in this paper.
The lack of experiments on slender poroelastic filaments being deformed on the 
poroelastic time scale prevents us from testing our predictions quantitatively. 
Since our results are relevant to swollen polymer networks, gel actuators and sensors, the mechanics
of cartilaginous joints, and the physics of rapid movements in plants, an important next step is the quantitative experimental study of slender poroelastic structures.

\begin{table}
\label{tab2}
\begin{tabular}{|l|l|l|l|l|l|} \hline
Application & $\mu$ ($Pa$)  & $\alpha$ & $k$ ($m^2/Pa\,sec$)& $R$ ($m$) & $\tau_p$ ($sec$) \\ \hline \hline
Actin cytoskeleton & $100$ & 0.8 & $10^{-12}$ & $10^{-6}$ & $10^{-2}$ \\ \hline
Bones & $10^{10}$ & 0.05 & $10^{-14}-10^{-16}$ & $10^{-2}$ & $0.1 - 10^{-3}$ \\ \hline
Cartilage  & $10^6$ & 0.8 & $(1-6)10^{-16}$ & $10^{-3}$ & $10^3$ \\ \hline
Plant stem/root & $10^8$ & 0.8 & $10^{-11}$ & $10^{-2}$ & $10^{-2}$ \\ \hline
Venus' fly trap leaf & $10^6$ & 0.8 & $10^{-12}$ & $10^{-3}$ & 0.1 \\ \hline
%
%
\end{tabular}
\caption{Applications of poroelasticity in biology.}
\end{table}

\section*{Acknowledgments} We acknowledge support via the Norwegian Research Council (JS), the US Office of Naval Research Young Investigator Program (LM),  the US National Institutes of Health (LM) and the Schlumberger Chair Fund (LM).  The authors thank Mederic Argentina for insightful discussions.

\appendix*{Appendix A : Derivation of the poroelasticity equations}
\label{appendix}

\renewcommand{\theequation}{A-\arabic{equation}}

The derivations of the equations of poroelasticity have been many and varied.  
Partly this has been because several qualitatively different parameter regimes 
containing distinct leading order force balances exist.  
We focus here on the equations which govern the 
second row of Table 1, namely where the Stokes' length is much larger than the pore size, 
{\it i.e.} L$_s \gg l_p$.  
The methods used to derive equations for this region of parameter space can be
 classified into three categories: physical arguments and superposition (Biot 1941, Biot \& Willis 1957),
 mixture theory (Barry \& Holmes 2001), and micro structural derivations 
(Auriault \& Sanchez-Palencia 1977, Burridge \& Keller 1981, Mei \& Auriault 1989). 
 First, we show in detail a version of the micro structural derivations, which uses ideas 
from both Burridge \& Keller (1981) and Mei \& Auriault (1989).

\begin{figure}[h]
\begin{center}
\includegraphics[width=7cm]{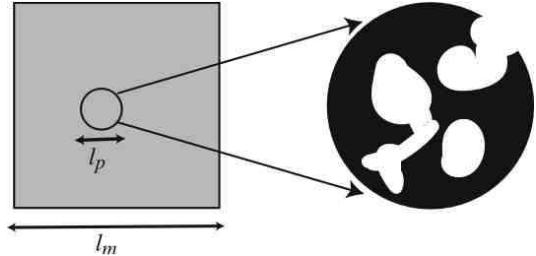}
\par
\caption{Typical porous medium illustrating the separation of length scales, where $l_p$ is the pore scale and $l_m$ is the system scale.}
\label{pe}
\end{center}
\end{figure}

The equations that govern the behavior in the incompressible interstitial fluid at low Re are
\begin{eqnarray}
\label{fs}
\boldsymbol{\sigma}_f = -p{\bf I} + 2\epsilon\mu{\bf e (v)}, \\
\nabla\cdot\boldsymbol{\sigma}_f = 0, \\
\nabla\cdot {\bf v} = 0,
\end{eqnarray}
where $\sigma_f$ is the stress tensor in the fluid, $\epsilon = l_p/l_m$ (see figure \ref{pe}), 
${\bf e}(..) = \frac{1}{2}[\nabla(..) + \nabla(..)^T]$ is the strain operator, and ${\bf v}$ is the fluid velocity.
In the solid the analogous equations are
\begin{eqnarray}
\label{ss}
\boldsymbol{\sigma}_s = {\bf A}:{\bf e (u)}, \\
\label{fsb}
\nabla \cdot \boldsymbol{\sigma}_s = 0,
\end{eqnarray}
where ${\bf u}$ is the displacement field, ${\bf A}$ is the tensor of elastic moduli,
and $\sigma_s$ is the stress tensor in the solid. 
At the solid-fluid interface continuity of displacements and tractions yields 
\begin{eqnarray}
\label{bc2}
{\bf v - \partial_tu} = 0, \\
\label{bc1}
\boldsymbol{\sigma}_s\cdot{\bf n} - \boldsymbol{\sigma}_f\cdot{\bf n} = 0.
\end{eqnarray}
Here ${\bf n}$ is the unit normal vector to the surface separating the two phases.

Looking for a perturbation solution in powers of the small parameter $\epsilon$, we use
an asymptotic expansion of the variables
\begin{eqnarray}
\boldsymbol{\sigma_f} = \boldsymbol{\sigma}_f^0 + \epsilon\boldsymbol{\sigma}_f^1 + ... \nonumber \\
\boldsymbol{\sigma}_s = \boldsymbol{\sigma}_s^0 + \epsilon\boldsymbol{\sigma}_s^1 + ... \nonumber \\
p = p^0 + \epsilon p^1 + ... \nonumber \\
{\bf u} = {\bf u}^0 + \epsilon{\bf u}^1 + ... \nonumber \\
{\bf v} = {\bf v}^0 + \epsilon{\bf v}^1 + ...
\end{eqnarray}
with a multiple-scale expansion for the gradient 
\begin{equation}
\nabla = \nabla_{x'} + \epsilon\nabla,
\end{equation}
where ${\bf x}$ denotes the macroscopic scale, ${\bf x'} = \epsilon {\bf x}$ denotes the pore scale, 
$\nabla$ denotes the gradient relative to the macroscopic scale and 
$\nabla_{x'}$ denotes the gradient relative to the pore scale.  
Since we assume that the flow is driven on the macroscopic scale, the leading order deformation is a function only of ${\bf x}$. Then equations (\ref{fs}) and (\ref{ss}) yield the following
expressions for the fluid and solid stress tensors:
\begin{eqnarray}
\boldsymbol{\sigma}_s^0 = {\bf A}:[{\bf e (u}^0) + {\bf e}_{x'} ({\bf u}^1)], \\
\boldsymbol{\sigma}_s^1 = {\bf A}:[{\bf e (u}^1) + {\bf e}_{x'} ({\bf u}^2)], \\
\boldsymbol{\sigma}_f^0 = -p^0{\bf I}, \\
\boldsymbol{\sigma}_f^1 = -p^1{\bf I} + \mu{\bf e}_{x'}({\bf v}^0), 
\end{eqnarray}
where ${\bf e}$ and ${\bf e_{x'}}$ denote the strain relative to the system scale and pore scale coordinates respectively.  The stress balance in the fluid (\ref{fsb}) yields:
\begin{eqnarray}
\nabla_{x'}p^0=0, \\
\label{bal}
\mu\nabla^2{\bf v}^0 - \nabla_{x'}p^1 - \nabla p^0 = 0.
\end{eqnarray}
Thus the leading order pressure gradient $p^0({\bf x})$ is only a function of the system scale coordinate.
The stress balance in the solid yields
\begin{eqnarray}
\nabla_{x'}\cdot\boldsymbol{\sigma}_s^0 = 0, \\
\nabla_{x'}\cdot\boldsymbol{\sigma}_s^1 + \nabla\cdot\boldsymbol{\sigma}_s^0=0.
\end{eqnarray}
we define $\boldsymbol{\sigma}$ to be the total stress tensor:
\begin{eqnarray}
\boldsymbol{\sigma} = \boldsymbol{\sigma}_s ~~ \hbox{in} ~~ V_s, \nonumber \\
\boldsymbol{\sigma} = \boldsymbol{\sigma}_f ~~ \hbox{in} ~~ V_f,
\end{eqnarray}
where $V_s$ and $V_f$ are the solid and fluid parts of a volume element.  Stress balance in the fluid and solid imply 
\begin{equation}
\label{st}
\nabla_{x'}\cdot\boldsymbol{\sigma}^1 + \nabla\cdot\boldsymbol{\sigma}^0 = 0.
\end{equation}
Averaging (\ref{st}) over the pore scale.
\begin{equation}
\frac{1}{V}\int\nabla\cdot\boldsymbol{\sigma}^0dV + \frac{1}{V}\int\nabla_{x'}\cdot\boldsymbol{\sigma}^1dV = \frac{1}{V}\int\nabla\cdot\boldsymbol{\sigma}^0dV +\frac{1}{V}\int{\bf n}\cdot\boldsymbol{\sigma}^1dS = 0,
\end{equation}
where $V = V_f + V_s$.  In the limit $V\to\infty$, $\frac{1}{V}\int{\bf n}\cdot\boldsymbol{\sigma}^1dS \to 0$ 
since the surface to volume ratio tends to zero.  Consequently,
\begin{eqnarray}
\frac{1}{V}\int\nabla\cdot\boldsymbol{\sigma}^0dV = \nabla\cdot<\boldsymbol{\sigma}^0> = 0, \\
<\boldsymbol{\sigma}^0> = <{\bf A}:[{\bf e (u}^0) + {\bf e}_{x'} ({\bf u}^1)]> - \phi_fp^0{\bf I},
\end{eqnarray}
where $\phi_f$ is the fluid volume fraction and $< >$ denotes averages over the pore scale.  
In order to write the averaged equations in terms of ${\bf u}^0$ and $p^0$,  we must 
eliminate ${\bf u}^1$.  This is achieved by using the stress balance in the solid so that
\begin{equation}
\label{ms1}
\nabla_{x'}\cdot \sigma_s^0 = \nabla_{x'}\cdot \{ {\bf A}:[{\bf e} ({\bf u}^0) + {\bf e}_{x'} ({\bf u}^1)] \} = 0.
\end{equation}
The boundary condition (\ref{bc1}) at the fluid solid surface yields
\begin{equation}
\label{ms}
{\bf A}:[{\bf e (u}^0) + {\bf e}_{x'} ({\bf u}^1)]\cdot {\bf n} = -p^0 {\bf n}.
\end{equation}
Since this is a linear system of equations, ${\bf u}^1$ is a linear combination of $p^0$ and ${\bf e (u}^0)$:
\begin{equation}
\label{u1}
{\bf u^1} = {\bf B}:{\bf e (u}^0) - {\bf C}p^0,
\end{equation}
where the third rank tensor ${\bf B}$ and vector ${\bf C}$ vary on the pore and system scales, 
and can only be found explicitly by solving the micro structural problem (\ref{ms1})-(\ref{ms}).  
The averaged stress tensor becomes
\begin{equation}
\label{sig}
<\boldsymbol{\sigma}^0> = <{\bf A} + {\bf A}:{\bf e_{x'}(B)} >:{\bf e(u}^0) - <{\bf A}: {\bf e_{x'}(C})>p^0- \phi_f p^0 {\bf I},
\end{equation}
where in index notation 
${\bf e_{x'}(B)}=(\partial_{x'_m}B_{nkl} + \partial_{x'_n}B_{mkl})$
If we assume that the material is isotropic on the macroscopic scale we can further reduce (\ref{sig}):
\begin{equation}
\label{pe1}
<\sigma^0> = 2\mu{\bf e (u}^0) + \lambda\nabla\cdot{\bf u}^0{\bf I} + (-\phi_f + \gamma)p^0{\bf I},
\end{equation}
where $\gamma{\bf I} = <{\bf A:e(C)}>$ is an isotropic pressure in the solid due to the fluid pressure exerted at the interface.
Substituting (\ref{pe1}) into the stress balance equation
\begin{equation}
\label{pe3}
\nabla\cdot<\boldsymbol{\sigma}^0> = 0,
\end{equation}
gives us three equations for the four unknowns (${\bf u}^0$ and $p^0$).  
We now turn to continuity to give us the final equation.
Continuity will give the final equation.  Since the fluid stress balance (\ref{bal}) is linearly forced by the external 
pressure gradient we can define a tensor ${\bf k}$ relating the external pressure gradient to the pore scale flow:
\begin{equation}
\label{darcy}
{\bf v}^0  - \partial_t{\bf u}^0 = -{\bf k}\cdot\nabla p^0,
\end{equation}
Averaging over the fluid volume yields
\begin{equation}
<{\bf v}^0> - \phi_f\partial_t{\bf u}^0 = -<{\bf k}>\cdot\nabla p^0.
\end{equation}
Since the fluid is incompressible averaging the continuity equation 
\begin{equation}
\nabla\cdot{\bf v}^0 + \nabla_{x'}\cdot{\bf v}^1=0,
\end{equation}
gives
\begin{eqnarray}
0  = \nabla\cdot<{\bf v}^0> + \frac{1}{V}\int\nabla_{x'}\cdot {\bf v}^1dV = \nabla\cdot<{\bf v}^0> + \frac{1}{V}\int{\bf n}\cdot {\bf v}^1dS 
 \nonumber \\ 
\label{v0}
= \nabla\cdot<{\bf v}^0> + \frac{1}{V}\int{\bf n}\cdot \partial_t{\bf u}^1dS
 = \nabla\cdot<{\bf v}^0> - \frac{1}{V}\int\nabla\cdot  \partial_t{\bf u}^1dV,
\end{eqnarray}
where we have used (\ref{bc2}).  Taking the divergence of (\ref{darcy}) and using (\ref{v0}) and (\ref{u1}) to eliminate ${\bf v}^0$ and ${\bf u}^1$ respectively yields
\begin{eqnarray}
-\nabla\cdot<{\bf k}>\cdot\nabla p^0 = \nabla\cdot(<{\bf v}^0>  - \phi_f\partial_t{\bf u}^0) \nonumber \\
\label{rhs}
= <\nabla_{x'}\cdot {\bf B}>:{\bf e}(\partial_t{\bf u}^0) - <\nabla_{x'}\cdot{\bf C}>\partial_tp^0 - \partial_t{\bf u}^0 \cdot\nabla\phi_f - \phi_f\nabla\cdot\partial_t{\bf u}^0.
\end{eqnarray}
If the change in solid volume fraction is much smaller than the volume fraction itself $\partial_t {\bf u} 
\cdot \nabla \phi_f \approx 0$.  Furthermore, if the solid skeleton 
is incompressible then $\nabla\cdot<{\bf v}^0> = 0$ so that the first two terms on the right hand side of equation (\ref{rhs}) are negligible.  For a compressible isotropic skeleton (\ref{rhs}) yields 
\begin{equation}
\label{pe2}
\beta \partial_t p^0 - \nabla \cdot <{\bf k}>\cdot\nabla p^0 = -\alpha\partial_t\nabla\cdot{\bf u}^0,
\end{equation}
where the $\beta = <\nabla_{x'}\cdot{\bf C}>$ is the bulk compliance of the solid skeleton and $\alpha= \phi_f - <\nabla_{x'}\cdot {\bf B}>_{ii}/3$ is the effective fluid volume fraction. In general if the solid is treated as compressible, the fluid must also be treated as such since their bulk moduli are comparable. Thus, $\beta$ is really a measure of the compressibility when the system is jacketed, so that for a mixture of an incompressible solid and fluid, $\beta=0$. Multiple scale analysis (Auriault \& Sanchez-Palencia 1977) shows that $<\nabla_{x'}\cdot {\bf B}> =  <{\bf A:e(C)}>=\gamma$ so that (A27) takes the final form
\begin{equation}
\label{lpe}
<\boldsymbol{\sigma}^0> = 2\mu{\bf e(u}^0) + \lambda \nabla \cdot {\bf u}^0\,{\bf I} 
- \alpha p^0{\bf I}.
\end{equation}
Equations (\ref{pe3}), (\ref{pe2})  and (\ref{lpe}) are the equations of poroelasticity, identical in form to the equations written down by Biot (1941).  Removing the brackets and superscripts we recover equations (\ref{CE}) and (\ref{Cont}) from \S \ref{goveq}.  Studying poroelasticity from the micro structural point of view allows us to see that Biot's (1941)
equations correspond to a locally compressible solid skeleton and the equations of mixture theory (Barry \& Holmes  2001) 
correspond to an incompressible solid skeleton.

\appendix* {Appendix B: Equations of motion for a poroelastic sheet}
\label{plate}

\renewcommand{\theequation}{B-\arabic{equation}}

The  equations of motion for a sheet of thickness $H$ and length $L$   are found using the same techniques as for a filament. The displacement field ${\bf u}=(u(y),v(y),0)$ is two dimensional, where the $y$ direction is normal to the neutral surface and the free surfaces are located at $y=\pm H/2$. We use the following dimensionless parameters 
\begin{eqnarray}
 t = (\beta + \frac{\alpha^2}{2\mu + \lambda})\frac{H^2}{k} \,t', ~~~~~~ p= \frac{2\mu\alpha}{[\beta(2\mu+\lambda)+\alpha^2]}\frac{H^2}{L^2}\,p',  \nonumber \\
y=H\, y', ~~~ \sigma_{xx} = \frac{4\mu(\mu+\lambda)}{2\mu+\lambda}\frac{H^2}{L^2}\, \sigma_{xx}',  ~~~
P = \frac{\mu(\mu+\lambda)H^3}{3(2\mu+\lambda)L^2}\,P'.
\end{eqnarray}
The dimensionless parameter $\delta$ characterising the ratio of the fluid dtress to the solid stress is
\begin{equation}
\delta = \frac{12\mu\alpha^2}{(\mu+\lambda)[\beta(2\mu+\lambda)+\alpha^2]}
\end{equation}
The pressure field is found by solving the 1-dimensional diffusion equation 
\begin{equation}
\partial_tp - \partial_{yy}p = -y\,\partial_{xt}\theta,
\end{equation}
with the boundary conditions
\begin{eqnarray}
 \partial_y p + \hbox{Bi}\,p = 0 ~~ \hbox{at} ~~ y=  1/2, \nonumber \\
-\partial_y p + \hbox{Bi}\,p = 0 ~~ \hbox{at} ~~ y=-1/2. 
\end{eqnarray}
Then
\begin{equation}
p=-\sum_n\chi_n\,\sin \sqrt{\lambda_n}y\,\int_0^te^{-\lambda_n(t-t')}\partial_{xt'}\theta\,dt'
\end{equation}
where the $\lambda_n$ satisfy
\begin{equation}
\label{B6}
\sqrt{\lambda_n} \cos\frac{\sqrt{\lambda_n}}{2} + Bi\,\sin\frac{\sqrt{\lambda_n}}{2}=0,
\end{equation}
and $\chi_n$ and $\gamma_n$ are given by
\begin{equation}
\label{B7}
\chi_n=\frac{2(2+Bi)\sin\frac{\sqrt{\lambda_n}}{2}}{\lambda_n(1-\frac{\sin\sqrt{\lambda_n}}{\sqrt{\lambda_n}})},~~~
\gamma_n=\frac{2(2+Bi)^2\sin^2\frac{\sqrt{\lambda_n}}{2}}{\lambda_n^2(1-\frac{\sin\sqrt{\lambda_n}}{\sqrt{\lambda_n}})}.
\end{equation}
Equations (\ref{B6}) and (\ref{B7}) together with equation (\ref{poro}) for the motion of a poroelastic plate with time-dependent plane stress.

\appendix*{Appendix C: Kirchhoff-Love theory for a bent, twisted filament}
\label{5}

\renewcommand{\theequation}{C-\arabic{equation}}

In this section we construct the equilibrium equations for a thin poroelastic rod whose deformation is not necessarily in the plane. The case of a purely elastic filament is treated in Love (1944). The configuration is given by the position of the centre line and the orientation of its cross-section at every point along it.  At every point along the centre line of the rod ${\bf r(X,t)} = (X(x,t),Y(x,t),Z(x,t))$, where $x$ is the arc-length we consider the orthogonal triad ${\bf d}_i(x,t)$, $i=1,2,3,$ where ${\bf d}_1$ and ${\bf d}_2$ lie 
along the principal axes of the cross-section of the rod and 
\begin{equation}
\label{kine}
{\bf d}_3 = \partial_x{\bf r}
\end{equation}
is the vector tangent to the centre line.  The orientation is determined by a body-fixed director frame that  allows us to consider finite deformations. In this case we use the director basis to follow the evolution of the fluid pressure field.
The vector of strains \boldmath $ \kappa$ \unboldmath is given by
\begin{equation}
\hbox{\boldmath$\kappa$\unboldmath} = \kappa^{(1)}{\bf d}_1  +\kappa^{(2)}{\bf d}_2 + \Omega {\bf d}_3,
\end{equation}
which defines the rotation of the principal axes along the filament. 
Here $\kappa^{(1)}$ and $\kappa^{(2)}$ are the projections of the curvature of the centre-line onto the principal axes of the cross-section 
and $\Omega$ is the twist strain.
\begin{equation}
\partial_x {\bf d}_i = \hbox{\boldmath$\kappa$\unboldmath} \times {\bf d}_i.
\end{equation}
The stress resultant vector ${\bf F}(x,t)$ and the couple resultant vector ${\bf M}(x,t)$ at any cross-section can be written as 
\begin{eqnarray}
{\bf F} = \sum_{i=1}^3F^{(i)}(x,t){\bf d}_i(x,t), ~~~ {\bf M} = \sum_{i=1}^3M^{(i)}(x,t){\bf d}_i(x,t),
\end{eqnarray}
where $F^{(1)}$ and $F^{(2)}$ are the shear forces and $M^{(1)}$ and $M^{(2)}$ are the bending moments along the principal axes, $F^{(3)}$ 
is the tensile force and $M^{(3)}$ is the twisting moment.  

Since the equation for the diffusion of pressure is linear we 
consider the bending about the principal axes separately.  In light of equation (\ref{fmom}) above ($\partial_x\theta$ being the curvature along one
of the principal axes) we can write equations for the dimensionless couple resultant vector ${\bf M}$
\begin{eqnarray}
\label{MM}
M^{(1)} = \kappa^{(1)} + \delta \sum_{n=1}^\infty \gamma_n \int_0^te^{-\lambda_n(t-t')}\partial_{t'}\kappa^{(1)} dt', \nonumber \\
M^{(2)} = \kappa^{(2)} + \delta \sum_{n=1}^\infty \gamma_n \int_0^te^{-\lambda_n(t-t')}\partial_{t'}\kappa^{(2)} dt', \nonumber \\
M^{(3)} = C\,\Omega,
\end{eqnarray}
where $C(=\frac{2(\lambda+\mu)}{3\lambda+2\mu}$for a circular rod) is the dimensionless torsional rigidity 
(normalised by the bending contribution to $M^{(1)}$), 
 $\tau_a$ is the dimensionless twist strain,
and the $\lambda_n$ are determined by solving equation (\ref{lam}).
We note that the twisting moment has no poroelastic contribution because it is purely a shear deformation, and poroelastic 
effects arise only from volumetric deformations as seen in equation (\ref{Cont}). Finally, the local balance of forces and torques give the equilibrium
equations
\begin{eqnarray}
\label{FFF}
\partial_x{\bf F} + {\bf F}_{external} = 0, \\
\label{MMM}
\partial_x{\bf M} + {\bf d}_3 \times {\bf F} = 0,
\end{eqnarray}
where ${\bf F}_{external}$ is the external body force acting on the cross-section.  The complete set of equations
that determine the poroelastic behaviour of a filament are (\ref{kine}),(\ref{MM}),(\ref{FFF}) and (\ref{MMM}).


\label{lastpage}

\end{document}